\documentstyle[sprocl]{article}
\def\be{\begin{equation}}
\def\ee{\end{equation}}
\def\bea{\begin{eqnarray}}
\def\eea{\end{eqnarray}}

\begin{document}

\title{
\begin{flushright} \small
   DTP--MSU/01-21\\
   hep-th/0112038
  \end{flushright}
\vspace{1cm} \textbf{GRAVITATING LUMPS}   }
\author{D. V. GAL'TSOV}

\address{
Department of Theoretical Physics,\\
    \sl Moscow State University, 119899, Moscow, Russia\\
 Email: galtsov@grg.phys.msu.su}

\maketitle\abstracts{ Recent progress in the study of solitons
and black holes in non-Abelian field theories coupled to gravity
is reviewed. New topics include gravitational binding of
monopoles, black holes with non-trivial topology, Lue-Weinberg
bifurcation, asymptotically AdS lumps, solutions to the
Freedman-Schwarz model with applications to holography,
non-Abelian Born-Infeld solutions.}

\section{Introduction}
During the past decade  many  surprising phenomena were discovered
in the study of gravitating solitons and black holes involving
Yang-Mills (YM) fields. A key role in this development has been
played by gravitational sphalerons, known as Bartnik-McKinnon
(BK) particle-like solutions to the Einstein-Yang-Mills (EYM)
equations. The associated black holes, violating no-hair   and
uniqueness theorems, attracted much attention and helped to
clarify a number of beliefs persistent in the Einstein theory.
Other conceptually important results were obtained in the study
of self-gravitating magnetic monopoles in gauge theories with
spontaneous symmetry breaking. Meanwhile, for many years  this
research  basically served only as a theoretical laboratory to
study gravity in unusual conditions. Direct physical applications
were unknown until the recent progress in superstring theory
culminated in the discovery of D-branes and the related gauge
theory/supergravity correspondence. It turned out that some
gravitating lumps have non-trivial ten-dimensional interpretation
and are extremely useful in the holographic treatment of the
${\cal N}=1$ supersymmetric gauge theory. The goal of this
talk\footnote{This is a written version of the talk given at the
16th International
 Conference on General Relativity and Gravitation, held on July 15-21,
 2001, in Durban, South Africa} is to give a brief overview of  the subject
with an emphasis on the results obtained during the last three
years. A review of the work done before 1998 was given in
\cite{Volkov:1998cc} where we address the reader for references
prior to November 1998, citing only names and years.
\section{Gravitational sphalerons}
 Neither pure gravity, nor flat space Yang-Mills (YM) theory admit
static particle-like solutions. Gravity is  purely attractive, and
no static equilibrium is possible, unless some repulsive matter
is introduced. On the contrary, the YM field field is repulsive:
due to scale invariance, the trace of the energy momentum tensor
is zero, so the positivity of the energy density implies the
positivity of the sum of principal pressures. Gravity breaks
scale invariance, therefore the coupled Einstein-Yang-Mills (EYM)
theory  is expected to  contain particle-like solutions.
  There is also  a
topological argument similar to Manton-Taub explanation of the
existence of the sphaleron  in the Weinberg-Salam theory. In that
theory  the Higgs broken phase manifold is $S^3$. Non--triviality
of the  third homotopy group $\pi_3 (S^3)$ is an indication of the
presence of non--contractible loops in the configuration space
for asymptotically flat solutions, and, hence, saddle points on
the energy surface. The corresponding unstable solution is {\em
the} sphaleron; it sits on the top of the potential barrier
separating topologically distinct Yang--Mills vacua. It was
argued (Gal'tsov and Volkov (1991), Sudarsky and Wald (1992))
that the same argument can be applied to the gauge group itself,
in fact $SU(2)\sim S^3$, once Higgs is replaced by another
attractive field of non-topological nature, like gravity or
dilaton.

Historically, particle-like  SU(2) EYM solutions were discovered
numerically by Bartnik and McKinnon (1988)   before this sphaleron
interpretation was suggested. The metric is spherically symmetric
and is parameterized by two functions $N,\,\sigma$ of a radial
variable: \be\label{ds}
ds^2=N\,\sigma^2\,dt^2\,-\,\frac{dr^2}{N}\,-\,r^2d\Omega, \ee
while the YM field is purely magnetic:
$A=(1-w)\left(T_\theta \sin\theta d\varphi-T_\varphi
d\theta\right),$
where $w$ is a real function of $r$ and $T_\theta,\,T_\varphi$
are spherical projections of the SU(2) generators.  This is a
configuration with the unit winding number. Higher winding
numbers lead to non-spherical solutions. An embedded Abelian
magnetic monopole with unit magnetic charge is $w\equiv 0$, while
$w=\pm 1$ correspond to the neighboring topologically distinct YM
vacua. Particle-like solutions are characterized   by regularity
at the origin \be \label{wex} w=1-br^2+O(r^4),\quad N=1-4b^2r^2
+O(r^4), \ee and asymptotic flatness: $N\to 1-2M/r$ with finite
mass $M$, together with $\sigma\to 1$. This implies that the YM
field asymptotically stays in one of the vacua \be \label{waex}
w=\pm(1-a/r)+O(1/r^2). \ee Parameters $b,\,M,\,a$ can be
determined via numerical matching. Solutions exist for a discrete
increasing sequence $b_n$ on the semi-open interval
$[\,b_1=0.4537,\, b_\infty =0.706)$, with masses $M_n$ converging
to unity as $n\to\infty$ in the units $m_{Pl}/g$, where $g$ is the
gauge coupling constant. The function $w_n$ has $n$ zeroes and
tends to $(-1)^n$, solutions exist for any finite $n$. The
limiting solution, corresponding to $b_\infty$, is asymptotically
non-flat.

For the lowest $n=1$ BK solution, $w$ is a decreasing function
 varying between $w=1$ and $w=-1$ in exactly the same way, as
in the electroweak sphaleron. Odd-$n$ solutions all have
sphaleron  nature, while even-$n$ ones are topologically trivial.
A vacuum to vacuum path was constructed explicitly (Gal'tsov and
Volkov (1991)), showing that the Chern-Simons number is $1/2$ for
odd-$n$ solutions and zero for even $n$. Sphaleron interpretation
is supported by the fact that the $n$-th solution has exactly $n$
{\em odd}-parity negative modes, corresponding to excitations in
the electric sector. In addition, there are also $n$ {\em
even}-parity negative modes exhibiting gravitational instability
of BK solutions. Spherically symmetric magnetic configurations of
the SU(N) YM field are described by $N-1$ real functions, the
corresponding BK-type solutions have a nodal structure determined
by the set of $N-1$ integers (Kleihaus, Kunz and Sood
(1995-1996)).

Magnetic configurations  with the winding number $\nu$ could be
obtained by replacing $\varphi\to \nu\varphi$ in the YM ansatz,
but from the EYM equations it follows that there are no $\nu\neq
1$ spherically symmetric solutions other than embedded Abelian.
Essentially non-Abelian configurations with $\nu\geq 2$ can be
axially symmetric, in which case purely magnetic field is
parameterized by four real functions subject to one gauge
condition. Static axially symmetric regular solutions of the BK
type were found numerically for lower $\nu=2,3,4$ (Kleihaus and
Kunz (1997)), they are likely to exist for all $\nu$. The SU(2)
solutions are characterized by a node number $n$, their masses
increase both with growing $\nu$ and $n$, approaching $M=\nu$ for
$n\to\infty$. The metric is parameterized by three functions
entering the Papapetrou ansatz;  they exhibit toroidal or
spheroidal shape of the equipotential surfaces. These solutions
can be interpreted as superposition of BK particles aligned along
$z$-axis, their Chern-Simons number is $\nu/2$.  It remains
unknown, whether static asymptotically flat EYM solutions exist
{\em without} axial symmetry, like multimonopoles with $\nu\geq
3$.

\section{Gravitating monopoles}
\subsection{Gravitational excitations}
Magnetic monopole in the Georgy-Glashow model possesses a mass
$M_{mon}\sim v/g$, where $v$ is the vacuum expectation value of
Higgs, and a radius $R_{mon}\sim 1/(gv)$. When the monopole
gravitational radius $R_G\sim G M_{mon}$ approaches $R_{mon}$,
one can expect new phenomena to occur. With this motivation, Lee,
Nair and Weinberg (1992), Ortiz (1992), Breitenlohner, Forgacs
and Maison (1992, 1995), see also \cite{Maison:2000hm}, have
investigated the coupled EYM-Higgs (EYMH) static spherically
symmetric system and   found that the existence of the maximal
value of the dimensionless parameter $\alpha=\sqrt{G} v$, which
is proportional to the ratio of the mass of $W$-boson in this
theory to the Planck's mass, such that self-gravitating monopoles
cease to exist for $\alpha>\alpha_{cr}$. This is not surprising,
since for $\alpha\sim 1$ the monopole radius and its
gravitational radius become comparable, so for large $\alpha$ the
monopole would be inside the event horizon.

An important new feature is the possibility of gravitational
excitation of  monopoles. Note that the YM function $w$ in the
monopole solution starts with $w=1$ at the origin and
monotonically tends to zero at infinity. There are other solutions
in which $w$ oscillates $n$ times around zero before to enter the
asymptotic regime, these correspond to higher values of the
parameter $b$ in (\ref{wex}), increasing with $n$, for each
$\alpha$. On the phase diagram $(b,\,\alpha)$ one observes the
lower branch extending up to $\alpha_{cr}$ (ground state
monopole) where it bifurcates giving rise to the upper branch
(excited monopole with one $w$-node). This upper branch
bifurcates in its turn giving rise to the next $n=2$ branch and
so on. All upper branches in the limit $\alpha\to 0$ tend to
BK$_n$ EYM solutions after a suitable rescaling. Solutions
belonging to the main branch are stable, like flat-space
monopoles, while BK-excited monopoles are unstable, like BK
particles. Excited monopoles can be regarded as bound states of
monopoles with BK particles inside. Solutions near the main
bifurcation point were recently investigated in   detail  by
 Lue and Weinberg~\cite{Lue:1999zp} revealing new surprising features.
 Other recent work on gravitating monopoles
includes the study of dyons~\cite{Brihaye:1999zt}, SU(3)
monopoles~\cite{Brihaye:2001td}, EYMH solutions with non-minimal
gravitational coupling \cite{VanderBij:2000cu}, YMH coupled to
Brans-Dicke gravity~\cite{Tamaki:1999er}.
\subsection{Gravitational binding of monopoles}
  Like
monopoles at rest in the flat space-time do not interact  in the
BPS limit where the Higgs field is massless, and  a scalar
attraction is precisely compensated by a magnetic repulsion. Away
from the BPS bound, the Higgs field is massive, and the scalar
force falls off exponentially, so magnetic repulsion dominates,
and the energy of multimonopole solutions per winding number is
always larger than the energy of the spherical $\nu=1$ monopole.
For $\nu\geq 3$ the BPS multimonopoles with no axial symmetry also
exist~\cite{Houghton:1996bs}.

It was found that gravity can bind monopoles provided the Higgs
self-interaction constant $\lambda$ is small enough. It was
verified~\cite{Hartmann:2000gx,Hartmann:2001ic} that the energy
per unit winding number for $\nu=2, 3$ aligned gravitating
monopoles  is smaller than the energy of a single $\nu=1$
monopole. In the case of monopole-antimonopole pairs,  gravity
enlarges the space of solutions: beside the ground state pair
solution, there are BK-excitations which bifurcate at  critical
$\alpha$. The upper branch in the  limit of vanishing $\alpha$
correspond to the (rescaled) BK$_1$ solution, which is
spherically symmetric. An excited monopole-antimonopole pair
linking to BK$_2$  was also found~\cite{Kleihaus:2000hx}, this is
conjectured to be valid for all $n$.

Other flat-space   particle-like solutions include sphalerons in
the Weinberg-Salam model and Skyrmions. Similarly to monopoles,
gravitating sphalerons exist   up to finite $\alpha$ and link  to
the BK-excitations (the bifurcation pattern is somewhat different
from that in monopoles, for details see~\cite{Volkov:1998cc}.
When Higgs self-interaction constant increases, new sphaleron
solutions, currently called {\em bi-sphalerons}, bifurcate  from
the main branch (Kunz and Brihaye (1989),Yaffe (1990)). Their
response to gravity was studied in~\cite{Brihaye:2000ez}. For
Skyrmions the self-gravitating problem was studied by Luckock
(1987) well before the corresponding monopole problem was
investigated. More detailed study was undertaken by Heusler, Droz
and Straumann (1991) and Bizon and Chmaj (1992). The moduli space
again exhibits bifurcations with  BK-excited branches.
\section{Hairy black holes}
\subsection{Violation of black hole rules}
Soon after discovery of BK solutions, it was shown that there also
exist black hole solutions with similar structure of fields
outside the event horizon. For them, instead of regularity
conditions (\ref{wex}) at the origin, the regularity of the event
horizon is required. The (non-degenerate) horizon is a linear
zero of $N:\, N(r_h)=0,\,N'(r_h)>0$, on which $w$ starts with a
finite value $w_h=w(r_h)$. This quantity serves as a parameter
labeling solutions, another parameter is the horizon radius
$r_h$. One finds black hole solutions for any $r_h$ and a
discrete sequence of $w_h$ within the interval $0<w_h<1$ (Volkov
and Gal'tsov (1989), Kunzle and Masood-ul-Alam (1990), Bizon
(1990)).

The EYM black holes violate a naive no-hair conjecture, since the
magnetic YM field outside the horizon is not associated with
conserved charges. Also, a uniqueness property is violated: an
infinite number of different solutions exists with equal ADM mass
and zero Coulomb charges. Black hole counterparts to axially
symmetric higher-$\nu$ solutions were found as well (Kleihaus and
Kunz, 1997), these obviously violate the Israel theorem, valid
for vacuum and electrovacuum, stating that static black holes
should be spherically symmetric. Further, the Birkhoff theorem is
no longer valid for the YM matter, since 'spin  from isospin'
phenomenon gives rise to field modes, which are spherically
symmetric, but time-dependent. Finally, the (electro)vacuum
staticity theorem does not hold (though  admits a non-Abelian
generalization), and the Einstein-Maxwell circularity theorem is
violated. All this stimulated a widespread interest to black
holes with YM hair as a laboratory to test various long-standing
beliefs of the black hole physics.

An internal structure of EYM black holes turned out to be
particularly interesting (Donets, Gal'tsov and Zotov (1996).
There are discrete sequences of black holes which have either
Schwarzschild or Reissner-Nordstr\"om (RN) type internal
structure, but a generic black hole exhibits a very unusual
interior. The metric function $N$ oscillates with an
exponentially growing amplitude remaining always negative, but
approaching zero closer and closer in subsequent cycles. Each
time when such an 'almost' Cauchy horizon is reached, the mass
function starts to inflate, making $N$ to turn back, so  true
Cauchy horizons never form. These oscillations never end, and the
approach to the singularity is non-analytic.  All solutions,
except for a discrete RN-type set (of zero measure in the moduli
space) satisfy the strong cosmic censorship hypothesis (no
internal Cauchy horizons).
\subsection{Monopole-black hole transition}
Gravitating Georgy-Glashow model solutions with
$\alpha>\alpha_{cr}$ contain an event horizon inside the monopole
core. This phenomenon is rather   generic (Kastor and Trashen
(1992)), it occurs each time when the lagrangian leads to the
relation $\varepsilon+p_r\sim N$ at a 'would be' event horizon
(the sum of the energy density and the radial pressure). It is so
for the EYMH systems with triplet or complex doublet Higgs and the
Skyrme model, all these theories admit hairy black holes. In
fact, the first such black hole was discovered in the Skyrme
model (Luckock and Moss (1986)), it is particularly interesting
as an example of a {\em stable} hairy black hole. Note, that
purely EYM black holes are unstable, $n$-node solution inherits
from  BK $2n$ negative modes. Skyrme black holes violate even a
weak form of the no-hair conjecture, ascribing the no-hair
property only to  stable black holes. For the recent work on
gravitating Skyrme model see~\cite{Moss:2000hf}. Dyonic EYMH
black holes were discussed in~\cite{Brihaye:1999nn}.

A monopole-black hole transition near the bifurcation threshold
was recently studied in detail by Lue and
Weinberg~\cite{Lue:1999zp}, see also a
review~\cite{Weinberg:2001gc}. Globally regular and black hole
space-times are qualitatively different, nevertheless, transition
between them passes through a continuous family of configurations.
Near the threshold $\alpha\sim\alpha_{cr}$ the metric function
$N$ develops a minimum at some point, attaining a value very
close to zero. This regular space-time has properties close to
those of black holes. One can associate with near-critical
solutions an {\em entropy}~\cite{Lue:2000nm}, since information
about the quasi-interior region is accessible with larger and
larger time delay as one approaches the critical point. This
entropy is likely to be compatible with the Bekenstein entropy.
Lue-Weinberg bifurcation was also discussed in other models
\cite{Brihaye:1999kt,Brihaye:1999zt}.
\subsection{Static non-spherical black holes}
Black hole counterparts to axially-symmetric multimonopole
solutions were constructed~\cite{Brihaye:1999kt,Hartmann:2001ic}
possessing a deformed event horizon with a constant surface
gravity. These disprove~\cite{Kleihaus:2001ti} a recent modified
uniqueness conjecture formulated in terms of the 'isolated
horizons' language~\cite{Corichi:2000dm}. Axial dyonic black
holes were constructed in~\cite{Hartmann:2000ja}. An interesting
case is the neutral black hole with a magnetic dipole
hair~\cite{Kleihaus:2000kv}. It is obtained by inserting a horizon
between two centers in the monopole-antimonopole solution. Black
holes in higher $\nu$ magnetic monopoles may be non-axisymmetric,
therefore violating Hawking's 'strong rigidity' (Ridgway and
Weinberg (1995)). Though no explicit solutions (even numerical)
were obtained so far, the argument showing their existence is
simple: there is an instability of the magnetically charged
(embedded Abelian) Reissner-Nordstr\"om solution, which may have
any magnetic charge, with respect to massive vector field present
in the spontaneously broken gauge theory, in  modes which are not
axially symmetric. Moreover, again due to 'spin from isospin',
there are {\em no} spherically symmetric states for $\nu\neq 1$.
Going beyond the linear approximation, one is able to show that
for $\nu=2$ the instability results in an axially symmetric
configuration, for $\nu=3$ --- tetrahedral, for $\nu=4$
--- cubic, etc. In fact, such crystal-type structures are known
in the flat space BPS multimonopole
systems~\cite{Houghton:1996bs}. Note that the internal structure
of hairy black holes with scalar fields is very different then
described above for the EYM case: near the singularity the scalar
field dominates and the mass-function exhibit a 'power-low
inflation' (Gal'tsov, Donets and Zotov (1977), Breitenlohner,
Lavrelashvili and Maison (1998)). For the Skyrme case
see~\cite{Tamaki:2001wc}.
\subsection{YM hair and quantum coherence}
String infinite symmetries may be responsible for the maintenance
of the quantum coherence in the black hole evaporation.  These
symmetries could encode all information about an enormous number
of massive string excitations accumulated in the black hole state.
Initially this idea was implemented using essentially
two-dimensional black holes (quantum W-hair). An attempt to
construct a four-dimensional model was made
in~\cite{Mavromatos:1999hp} invoking hairy black holes with the
$SU(\infty)$ YM field. The $su(N)$ algebra in the limit
$N\to\infty$ can be mapped to Poisson brackets defined on a unit
'internal' sphere. The static spherically symmetric ansatz,
elaborated by Kunzle (1991) for finite $N$, then gives rise to
the YM function depending on a radial variable and an internal
angle $\theta$. In presence of  the large negative cosmological
constant (with respect to the energy scale of the YM field) an
approximate solution was obtained describing a  black hole with
hair, depending on infinite number of parameters.
 \section{Asymptotically AdS lumps}
 \subsection{Nodeless solutions for EYM system}
Recent interest to AdS/CFT correspondence stimulated an
investigation of the EYM systems in presence of the negative
cosmological constant $\Lambda$. Hairy black holes in the SU(2)
theory were first studied in~\cite{Winstanley:1998sn}, and then
(together with regular solutions) in~\cite{Bjoraker:1999yd}.
Asymptotic conditions now  follow form a finiteness of the mass
$M$, which is a limiting value of the mass function $m(r)$
defined via $N=1-2m(r)/r -\Lambda r^2/3$ (also $\sigma\to 1$).
This no more implies $w\to \pm 1$, instead any constant value
$w_\infty$ is allowed. This means that solutions will have
non-zero magnetic charge $Q_m=(1-w_\infty^2)$. Since the
asymptotic condition on $w$ is relaxed, one finds solutions for
continuously varying parameter $b$ in (\ref{wex}) for each number
of nodes $n$ including zero. Topologically, the nodeless
solutions have no sphaleronic nature, so it can be expected that
they do not have odd parity negative modes. This was proved to be
the case both in the spherically symmetric
sector~\cite{Winstanley:1998sn,Bjoraker:1999yd,} and for
non-spherical perturbations~\cite{Sarbach:2001mc}.   The
gravitational (even parity) perturbations were studied in the
spherical sector in~\cite{Bjoraker:1999yd} and for non-spherical
excitations in~\cite{Sarbach:2001mc}. Spherical modes are stable
in the $n=0$ sector for $w_\infty <1/\sqrt{3}$. Another news is
that the 'non-Abelian baldness' (forbidding AF SU(2) dyons,
Gal'tsov and Ershov (1989,1990)) is no longer valid: there are
essentially non-Abelian solutions with an electric sector in the
SU(2) theory. One finds dyon solutions with finite $M$ for
continuously varying $b$ (and the corresponding parameter in the
electric sector) and any number of nodes $n$ including $n=0$.

Although these AdS monopoles and dyons are very different from
those in the AF EYMH theory, one finds again the bifurcation
patterns similar to described above:   branches with neighboring
$n$  get connected via bifurcation points in the parameter space.
In particular, the mass of nodeless monopole solutions as a
function of the magnetic charge exists  till some critical charge
value, then it turns back forming an upper unstable $n=1$ branch.
This feature, first observed in~\cite{Bjoraker:1999yd}, was
further studied in~\cite{Hosotani:2001iz}. Axially symmetric EYM
solutions with higher winding numbers  were investigated for
$\Lambda<0$ in a recent paper~\cite{Radu:2001ij}. Transition from
negative to positive $\Lambda$ apparently reveals   a fractal
structure of the moduli space~\cite{Bjoraker:1999yd}.
\subsection{Asymptotically AdS monopoles and sphalerons}
Solutions of the EYMH system with triplet Higgs were studied
in~\cite{Lugo:1999fm,Lugo:1999ai}. Contrary to the EYM case, no
essentially new features appear as compared with the $\Lambda=0$
theory. Like in the AF case, they exist in a limited region of
the parameter $Gv^2$, the critical value being somewhat smaller.
The theory with the doublet Higgs in presence of the negative
cosmological constant was investigated
in~\cite{VanderBij:2001ah}. Again, the situation is rather
similar to the $\Lambda=0$ case: the electric component of the YM
field is forbidden, no arbitrary $w_\infty $ is possible, a
critical value of the Newton constant is observed where solutions
cease to exist. However, non-zero $\Lambda$ implies a power law
fall-off of the fields at infinity. An extensive study of
perturbations of asymptotically AdS solitons and black holes was
performed recently~\cite{Sarbach:2001mc,Winstanley:2001bs}
confirming earlier results on stability of nodeless solutions.
\subsection{Black holes with non-spherical topology}
Once asymptotic flatness is abandoned, there is no more
'topological censorship', forbidding black holes with
non-spherical topology of the event horizon. Namely, for the
negative cosmological constant, two-dimensional section of the
horizon may be an arbitrary genus Riemann surface
\cite{Lemos:1995cq,Lemos:1996cz,Vanzo:1997gw}. Generalization of
topological censorship for higher genus black holes was discussed
in \cite{Woolgar:1999}. Asymptotically AdS solutions provide
non-trivial examples of black holes with the winding number one
which are topologically different  from the BK type solutions.
One has merely to replace the spherical element $d\Omega$ in
(\ref{ds}) by flat or hyperbolic
$d\Omega_k=d\theta^2+f_k(\theta)d\varphi^2$, with $k=1,\,0,\,-1$
and $f_1=\sin\theta,\,f_0=\theta,\,f_{-1}=\sinh\theta$.
 Purely magnetic nodeless $n=0$ black holes with
$k=-1,\,0$ were constructed numerically
in~\cite{VanderBij:2001ia}.   For $k=0$ the mass function remains
positive definite elsewhere, so the solutions have positive mass.
For the negative curvature $k=-1$, there are black holes not only
with positive, but also with zero or negative mass. Solutions
with $k=0$ are stable both in odd and even parity sectors, while
the hyperbolic solutions $k=-1$ are stable provided $w_\infty>1$.
Black hole solutions with non-spherical topology do not have
globally regular counterparts.
\section{Rotation}
\subsection{Failure of circularity}
 Mechanical equilibrium inside
the flat space static compact field configurations generically
will be destroyed once they are forced  to rotate, so a priori it
is unlikely that one could endow solitons with an arbitrary
angular momentum. On the other hand, in quantum theory, magnetic
monopoles generically acquire discrete angular momentum (spin)
due to interaction with fermions (fermion zero modes). This is
closely related to supersymmetric embeddings, so the possibility
of {\em proper} rotational  zero modes (involving bosonic fields)
is an important question. It is known that there are no such
modes satisfying Bogomol'nyi equation. More recently the absence
of non-Bogomol'nyi rotational zero modes       for EYMH
particle-like solutions was proved  by Brodbeck and Heusler
(1997) and Brodbeck, Heusler, Straumann and Volkov (1997). One
serious complication to study rotation non-perturbatively is a
failure of the circularity theorem for non-Abelian matter
(Heusler and Straumann, 1993). The problem is that the
Ricci-circularity condition for the stationary axially-symmetric
space-time is not ensured by just imposing the same symmetries on
the YM field. Therefore a Papapetrou ansatz does not necessarily
hold. The most general parameterizations was derived
in~\cite{Gal'tsov:1998af}. It is a 2D dilaton gravity, with the
matter sector including the 2D Yang-Mills field  , two 2D Higgs
fields  , as well as two scalar moduli  , the dilaton
 and two 2D Kaluza-Klein two-forms.
\subsection{Rotational zero modes}
Meanwhile, in the {\em linearized} theory, the number of
odd-parity zero modes involved in rotation is considerably less.
An analysis of perturbations of the EYMH system, extending
previous work on this subject (see~\cite{Volkov:1998cc} for a
review and references), was performed
in~\cite{Brodbeck:1999yc,Sarbach:1999gi}, showing, in particular,
stability of the Schwarzschild and Reissner-Nordstr\"om black
holes with respect to odd-parity EYM perturbations. It was
confirmed also that only $l=1$ zero modes, consisting of three
real-valued functions, are responsible for (infinitesimal)
rotation of both EYM and EYMH solitons and black holes. At this
level circularity {\em does} hold.  The set includes $h_{t
\varphi}$ metric perturbation and two isotopic variations of
$A_t$. They satisfy a coupled system of linear ordinary
differential equations of the second order, so the overall order
of the system is {\em six}. A necessary condition for  globally
defined continuously varying zero modes to exist is a non-trivial
intersection of the moduli spaces of local solutions near the
origin (the event horizon  black holes) and at spatial infinity.
A local analysis reveals that the moduli space at infinity is
{\em four}-dimensional for EYM (without Higgs) and {\em
three}-dimensional for EYMH. In both cases the dimension of the
moduli space near the origin (solitons) is {\em three}, while
that at the horizon is {\em four}. Thus, for solitons involving
scalar fields, the intersection of two three-dimensional moduli
spaces inside the full six-dimensional space is trivial. For
black holes inside magnetic monopoles, there is a one-parameter
intersection space, so event horizons may be rotating. For EYM
(BK) particle-like solutions the intersection space is also
one-dimensional, and more careful analysis shows that the
electric charge and angular momentum are excited simultaneously.
\subsection{Rotation and charge}
An essential difference between EYMH solitons and BK particles is
that the binding is non-gravitational in the first case and
gravitational in the second. So it is perhaps not surprising that
BK solutions possess rotational zero modes, while magnetic
monopoles do not. For black holes gravitational attraction is
dominating, so zero modes do exist in both cases. These may be
regarded as another admissible hair for static solutions.  Another
interesting feature is that the charge and rotational degrees of
freedom are connected in a different ways for different
solutions. Recall that the magnetic YM configuration in BK
particles is of the dipole type. An intuition based on the
properties of the Kerr-Newmann solution suggest to view the
magnetic moment of  charged rotating solutions as due to Faraday
effect. For rotating BK the situation is inverse: magnetic dipole
configuration is primary, while rotation leads to charging the
particle. For EYM black holes the angular momentum and the
electric charge are independent parameters, but the relationship
between the angular velocity of the event horizon and the angular
momentum defined at infinity is non-trivial. This is again
related to the 'spin from isospin' phenomenon:  there is an extra
contribution to the angular momentum from the black hole hair. In
particular, there exist black holes with zero angular velocity of
the horizon (static), but non-zero angular momentum (and non-zero
charge). Alternatively, there are black holes with rotating
horizons, but zero angular momentum, as seen from infinity. A
non-perturbative numerical analysis of rotating EYM black holes
with an assumption of circularity was reported
recently~\cite{Kleihaus:2000kg}, confirming consistency of this
assumption at non-linear level.
\section{Links to superstrings}
\subsection{Dilaton}
Since the YM field is scale-invariant, the only dimensional
parameter of the EYM theory is the Planck mass.   It is natural,
therefore, to look for physical applications in the superstring
theory. The first step to incorporate BK solutions into this
 framework consisted in including the dilaton.
Particle-like solutions similar to BK were found in the flat YMD
theory by Lavrelashvili and Maison (1992).  When gravity is also
taken into account, one finds a surprising equality of the ADM
mass and the dilaton charge of BK-type solutions (Donets and
Gal'tsov, 1993). This was shown to be is a consequence of the
scaling symmetry of the system of equation leading to the
relation $g_{tt}=\exp(2\phi)$ (Donets, Gal'tsov and Volkov,
1993). Another enigmatic feature was observed for the $n=1$ EYMD
solution: it turned out that the parameter $b$ in (\ref{wex}) is
rational, $b_1=1/6$, while $a_1=2M$ (Lavrelashvili and Maison,
1993). Efforts to explain these 'magic' numbers have led to a
fascinating development; before going to this, let us mention
some recent work on gravitating lumps with dilaton: axially
symmetric EYMD configurations~\cite{Kleihaus:1999ia},
Georgy-Glashow-dilaton monopoles~\cite{Forgacs:1998bp} and black
holes~\cite{Brihaye:2001bk}, dilaton binding in
multimonopoles~\cite{Brihaye:2001gg}.
\subsection{Freedman--Schwarz model}
In a search of hidden supersymmetry tentatively underlying these
features, Chamseddine and Volkov (1997) have investigated the
${\cal N}=4 \, SU(2)\times SU(2)$ gauged  supergravity also known
as Friedman-Schwarz (FS) model. Its bosonic sector includes two
YM fields, dilaton and axion (which can be set zero for static
purely magnetic configurations) with the dilaton potential,
proportional to $-(g_1^2+g_2^2)\exp(-2\phi)$. Formally one can
derive the SU(2) EYMD equations for $g_2=ig_1$. This trick does
not destroy supersymmetry: the supergravity Bogomol'nyi equations
survive. Later it was shown by Volkov~\cite{Volkov:1999et} that
such 'complexified' theory corresponds to the Euclidean FS model,
and the associated Bogomol'nyi equations admit indeed one-node
$w(r)$ as a solution~\cite{Volkov:1999vu,Volkov:1999jz} (though
still not found analytically). If one passes to the Lorentzian
metric in this solution (this does not return us to the initial
Lorentzian FS model), one obtains the EYMD analog of the $BK_1$,
which is, of course, not supersymmetric (and unstable), but still
bears an imprint of supersymmetry of its Euclidean counterpart.

Leaving only one SU(2) field in {\cal N=4} supergravity
('half-gauged'  FS model), one gets the lagrangian
$L=- R+2(\nabla\phi)^2-  {\rm e}^{2\phi}F^2/2+ {\rm e}^{-2\phi}$
which differs  from the simple EYMD one by the dilaton potential.
This theory admits an analytic
solution~\cite{Volkov:1997,Gubser:2001eg} 
\[ds^2=dt^2-dr^2-R^2d\Omega,\quad R^2=2r\coth(r+c)-w^2-1,\]
\be\label{cv}
w=r \sinh^{-1}(r+c),\quad   {\rm e}^{2\phi}=R^{-1}\sinh(r+c),
\ee
where the metric is presented in the string frame,
and $c$ is a real parameter.
The  solution  preserves $1/4$ of the initial supersymmetry of
the FS model
for all $c$.  Asymptotically $w\to 0$, so this is a charge one monopole.
However, the metric asymptotically is not flat and not AdS.
The $c=0$ solution is
globally regular,  an expansion of $w$ near the origin is of the type
(\ref{wex}) with 'magic' $b=1/6$. For $c\neq 0 $ ($c$ is assumed
to be non-negative),
the radius $R$ shrinks at some finite $r$, which is a curvature
singularity.
In the limit $c\to\infty$ one gets an Abelian dilatonic monopole.
\subsection{Holographic interpretation}
This solution has a ten-dimensional counterpart  in Type I (pure)
supergravity; in fact, as was shown by Volkov (1997), the $D=4$ FS
model corresponds to compactification of the latter on $S^3\times
S^3$. Somewhat unexpectedly, it was found~\cite{MaNu2000} that the
same solution describes wrapping of a NS5 brane of the IIB theory
on a shrinking $S^2$ . This opens a way to construct a holographic
description for the ${\cal N}=1$ super Yang-Mills. A particularly
interesting aspect of the latter theory, which can be
investigated using this duality, is the chiral symmetry breaking
transition. The non-Abelian component of the gauge field serves
as the order parameter: unbroken chiral symmetry corresponds to
having only Abelian YM component, $w=0$. Thus, the BK-type
solutions correspond to broken chiral symmetry, while a formation
of an Abelian black hole signals the symmetry restoration. So one
needs a detailed information about solutions including black
holes; this was the subject of the recent paper by Gubser,
Tseytlin and Volkov~\cite{Gubser:2001eg}. Other solutions than
(\ref{cv}) are non-BPS, they include globally regular solutions
which start at the origin with $b\neq 1/6$. Like for
asymptotically AdS BK solutions, the parameter $b$ here is
continuous, for $b<1/6$ $w$-solutions are still nodeless, while
for $1/6<b<1/2$, solutions with more and more nodes appear, with
the limiting solution $n=\infty$ starting with $b=1/2$. These
solutions may be regarded as BK-excited monopoles. This part of
the solution space is somewhat similar to solutions of the pure
EYM theory with a negative cosmological constant. However, now an
asymptotic value of $w$ is zero, indicating the charge one
monopole topology, while in the AdS case the magnetic (and
electric) charges are continuously varying. An interesting new
feature is that there is a discrete subspace of solutions with
$b=b_n$ possessing a finite energy. Apart  from these, there are
non-trivial solutions with $w\equiv\pm 1, 0$, which have
different ten-dimensional counterparts.

Black hole solutions are determined by the horizon value $w_h$.
For $w_h=0$ solutions are Abelian, while non-zero values $w_h$
give rise to non-Abelian black holes. These contain a discrete
subclass of finite-energy solutions for which $w$ falls off
exponentially at infinity. Black holes with toroidal event
horizon in the $U(1)\times U(1)$ sector of the FS model were
constructed in~\cite{Klemm:1998in}
\subsection{Gauss-Bonnet}
One earlier attempt to incorporate 'stringy' effects was to
include higher-curvature terms in the lagrangian. First $\alpha'$
correction to the heterotic string effective action includes the
Gauss-Bonnet term, $\beta \exp (2\phi) R_{GB},\; R_{GB}=
R_{\mu\nu\lambda\tau}R^{\mu\nu\lambda\tau}-4R_{\mu\nu}R^{\mu\nu}
+R^2$, which produces a non-trivial effect in four dimensions due
to the dilaton factor. Note by passing that this term does not
generate derivatives higher than the second order. Remarkably, in
the static spherical space-time this term does not destroy the
scale invariance of the equations of motion, so a non-trivial
first integral still exists (Donets and Gal'tsov, 1993) implying
the equality $M=D$. However, solutions exist only for restricted
values of the coefficient $\beta<\beta_n$, e.g. $n=1$ solution
disappears for $\beta_1=0.37$, the critical value decreases for
higher $n$. The question was raised whether  quadratic curvature
terms (Gauss-Bonnet) can  stabilize negative
modes~\cite{Kanti:1999sp}. It is known that Gauss-Bonnet acts as
a 'hair tonic' opening a channel of neutral black holes with a
dilaton charge (Kanti, Mavromatos, Rizos, Tamvakis and
Winstanley, 1996), and these are linearly stable in absence of
the vector fields (the same authors, 1998). Actually, in the
EYM(D) static spherically symmetric theory with the  Gauss-Bonnet
term there exists four types of black holes: i) neutral dilatonic
(YM sector non-excited), ii) usual BK type, iii) holes with the
unit magnetic charge (embedded Abelian, electric sector
non-excited) and vi) holes with an electric charge, also embedded
Abelian. It was found in~\cite{Kanti:1999sp} that those solutions
whose perturbations involve the YM magnetic function  (ii, iii)
remain unstable (in particular, sphaleronic instabilities of ii)
are untouched), while neutral and electric solutions are stable,
reflecting stability of Schwarzschild and Reissner-Nordstr\"om
black holes.

\section{Born-Infeld}
Another aspect related to superstrings consist in investigation
of solitons and black holes in the non-Abelian SU(N)
Dirac-Born-Infeld theory, arising as an effective theory on N
D3-branes~\cite{GiKu98}. If   branes are coincident, one deals
with the purely massless theory, while for separated branes there
is a Higgs potential with the vacuum expectation value equal to
the brane separation. The simplest case is the pure YM theory
with the
 Born-Infeld (BI) type action
\be\label{det} S=\frac{\beta^2}{4\pi}{\rm tr}\int
\left\{1-\sqrt{-\det(g_{\mu\nu}+\beta^{-1}F_{\mu\nu})}\right\}d^4
x \ee In the non-Abelian case, $F_{\mu\nu}$ is matrix-valued, and
one can a priori take the trace in the action in  different ways
(see~\cite{Dyadichev:2000iw} for details). Two definitions are
the most reliable: the symmetrized trace suggested by Tseytlin,
which well covers the lower orders of the perturbative string
effective action, and the 'ordinary trace-square root' form,
which is a direct non-Abelian generalization of the
four-dimensional U(1) BI action obtained by evaluating the
determinant under the square root. For the static spherically
symmetric  SU(2) magnetic ansatz   these definitions lead to
somewhat different one-dimensional lagrangians: \be\label{tr}
L_{tr}=\beta^2 r^2(1-{\cal} R)\quad {\cal R}
=(1+V^2+2K^2)^{1/2},\;\; \ee where $V^2=(1-w^2)^2/(2\beta^2
r^4),\;K^2=w'^2/(2\beta^2 r^2)$, and \be\label{str} L_{str}=
\beta^2 r^2\left[1-(1+V^2)^{1/2}+K^2{\cal
A}(1+V^2)^{-1/2}\right]\;\;\;\; \ee with ${\cal A}= W^{-1}\arctan
W,\; W=(1+V^2)^{1/2}(V^2-K^2)^{-1/2}$.
\subsection{NBI particle-like solutions (sphalerons)}
The BI lagrangian breaks scale invariance introducing a new new
scale parameter --- BI critical field $\beta$, which should be
regarded as a manifestation of non-locality of the underlying
string theory. Therefore, the no-go theorem for classical
glueballs in the flat space YM theory is overruled, and one is
led  to look for such solutions in the non-Abelian Born-Infeld
(NBI) theory.
 The proof of existence and numerical results for flat space NBI
particle-like solutions were presented in~\cite{Gal'tsov:1999vn}
for the ordinary trace (\ref{tr}) and in~\cite{Dyadichev:2000iw}
for the symmetrized trace (\ref{str}) lagrangians, they have
qualitatively similar properties. Solutions form a sequence of
the BK type,  but the difference is that now the parameter $b_n$
in the expansion (\ref{wex}) is rather large and rapidly growing
with the node number $n$, for example, $b_1=12.7,\, b_2=887$ in
the  case (\ref{tr}) (these values are even larger for the model
(\ref{str})~\cite{Dyadichev:2000iw}). Solutions have sphaleron
features and are expected to be unstable. When gravity is taken
into account \cite{Wirschins:2000ww,Dyadichev:2000xh}, one finds
that $b_n$, as a function of the dimensionless parameter
$G\beta$, continuously interpolates~\cite{Dyadichev:2000xh}
between these values  for small $G\beta$, and the BK values for
large $G\beta$.
\subsection{Damping of  gigantic oscillations inside YM black holes}
Black hole solutions in the Einstein-NBI theory were also
constructed~\cite{Wirschins:2000ww,Dyadichev:2000xh}, outside the
horizon  they are  qualitatively similar to those in the EYM case.
However, an {\em internal} structure of the ENBI black holes is
drastically different. In fact, one could expect that violent
oscillations inside the EYM black holes should be modified in
quantum theory: already in the first inflationary cycle the mass
function attains values exceeding the Planck  mass. Although the
ENBI theory is still classical, it incorporates string $\alpha'$
quantum corrections in a non-perturbative way, so one could
expect that the problem of the over-Planckian masses in the EYM
black holes will be resolved. This is indeed the
case~\cite{Dyadichev:2000xh}: the internal behavior of the mass
function now is perfectly smooth, though the singularity still
bears non-analytic features. Namely, the function $w$ attains a
final value $w_0$, and there exist a two-parameter family of
solutions for which the mass function is analytic in the
Schwarzschild-type singularity. This family, however, is not
generic, like in the EYM case. But now a generic three-parametric
solution  can be obtained in the vicinity of the singularity as
the series   expansion in terms of $r^{1/2}$. Mass function again
attains a finite value, but one finds that the singularity is
much weaker than in the Schwarzschild case. The scalar curvature
diverges only as $R\sim r^{-3/2}$. Note that non-analytic
behavior of the mass function $m(r)$ in the singularity is also
observed in the Abelian Einstein-Born-Infeld theory coupled to
dilaton~\cite{Clement:2000ue}.
\subsection{NBI monopoles}
Adding triplet Higgs to the NBI lagrangian, one finds flat space
monopoles~\cite{Grandi:1999dv,Grandi:1999rv} which exhibit
features, similar to those of gravitating monopoles. In view of
the existence of the flat-space NBI sphalerons, this is not
surprising. Monopole (and dyon) solutions exist for     $\beta$
varying  from infinity to a finite value $\beta_{cr}$, which was
shown~\cite{Galtsov:2000mk} to be a bifurcation point giving rise
to the branch of excited monopoles with one-node $w$ (higher node
excitations  exist as well). In~\cite{Galtsov:2000mk} a
non-commutative U(1) monopole (with Higgs) was also constructed
following the D-brane in B-field / non-commutative gauge theory
correspondence. Note that it is the symmetrized trace version of
the theory which leads to the BPS equations coincident with the
YMH ones. Effect of gravity on the NBI monopoles was discussed
in~\cite{Tripathy:1999zj,Tripathy:1999aw}. Recently the NBI model
with doublet Higgs was considered~\cite{Brihaye:2001af}, it was
found that the sphalerons in this theory  interpolate between the
usual electroweak sphaleron and the NBI sphalerons
of~\cite{Gal'tsov:1999vn}.
\section{Conclusion}
Basic properties of gravitating non-Abelian solitons and hairy
black holes are fairly well understood now and exhibit certain
universality:  character of gravitational excitations,
bifurcation patterns of moduli spaces, dependence of spectra on
asymptotic conditions, phenomena near 'almost' horizons. Our list
of references is  almost complete for the period 1999 -- November
2001 within the covered topics. However, many   related
directions remained outside the scope of the talk: global
monopoles, boson stars, solutions to the gravity coupled Abelian
Higgs model, cosmological solutions with non-Abelian fields,
topological inflation, critical collapse of the YM field,
solutions to higher-dimensional EYM systems. More complete update
to the review~\cite{Volkov:1998cc} will be given elsewhere.
\section*{Acknowledgments}
The author is grateful to LOC for hospitality during the conference and to
Gravity Research Foundation for support.
The work was also supported in part by the Russian Foundation for
Basic Research under grant 00-02-16306.
\section*{References}

\end{document}